\documentclass[final,3p,times,twocolumn]{article}
\usepackage{authblk}

\usepackage[colorinlistoftodos]{todonotes}

\usepackage{amsfonts}
\usepackage{amssymb}
\usepackage{amsmath}

\usepackage{booktabs}
\usepackage{caption}
\usepackage{csquotes}
\usepackage{float}
\usepackage{graphicx}
\usepackage{lineno}
\usepackage{multicol}
\usepackage{multirow}
\usepackage{subcaption}

\usepackage{algorithm}
\usepackage{algpseudocode}

\usepackage{pdflscape}

\newcommand{\hlc}[1]{{#1}}

\newtheorem{definition}{Definition}

\graphicspath{{figures/}}

\title{Foundations of Temporal Text Networks}

\author[]{Davide Vega}
\author[]{Matteo Magnani}
\affil[]{InfoLab\\Department of Information Technology\\
Uppsala University, Sweden\\
\{davide.vega, matteo.magnani\}@it.uu.se}

\begin{document}


\maketitle

\begin{abstract}
\small
Three fundamental elements to understand human information networks are the individuals \hlc{(actors)} in the network, the \hlc{information} they exchange, that is often \hlc{observable} online as text content 
(emails, social media posts, etc.), and the time when these \hlc{exchanges} happen. An extremely large amount of research has addressed some of these aspects either in isolation or as combinations of two of them. 
There are also more and more works studying systems where all 
three elements are present, but typically using ad hoc models and algorithms that cannot be easily transfered to other contexts. \hlc{To address this heterogeneity}, in this article we present a simple, expressive and extensible model for temporal text networks, that we claim can be used as a common ground across different types of networks and analysis tasks, and we show how simple procedures to produce views of the model allow the direct application of analysis methods already developed in other domains, from traditional data mining to multilayer network mining.

\end{abstract}

\section{Introduction}
\label{sec:intro}


A large amount of human-generated information is available online in the form of text exchanged between individuals at specific times.
Examples include social network sites, online forums and emails.
The public accessibility of several of these sources allows us
to observe our society at various scales, from focused conversations among small groups of individuals to broad political discussions involving heterogeneous audiences from large geographical areas~\cite{Hristova2017:Socialbehaviour,Nerghes2014:IntroPoliticalEconomical}.

This information is undoubtedly very valuable, as shown for example by the large revenues of big Internet companies and by its usage during political campaigns, but it is also very complex because of its joint textual, structural and temporal nature.
To cope with this complexity, researchers have typically focused on either the topology of the network, as commonly done in Network Science, or the text exchanged among individuals, using methods from Computational Linguistics.
In some cases time has also been taken into consideration as in, respectively, the fields of Temporal Networks and Temporal Information Retrieval.

However, despite this broad interest in human information networks, only a limited number of works have been developed to address text, network topology and time \hlc{in an integrated way and using a common data model}. In our opinion, this is partly a result of the over-specialization of today's academia, and the fragmented and discipline-specific development of network research. Unfortunately, omitting any of the three basic elements of temporal text networks may lead to significant information loss and prevent a deeper understanding of the information system, as exemplified in the next section.

\subsection{A motivating example}

\begin{figure*}[h!]
  \begin{subfigure}{0.3\textwidth}
    \includegraphics[width=\textwidth]{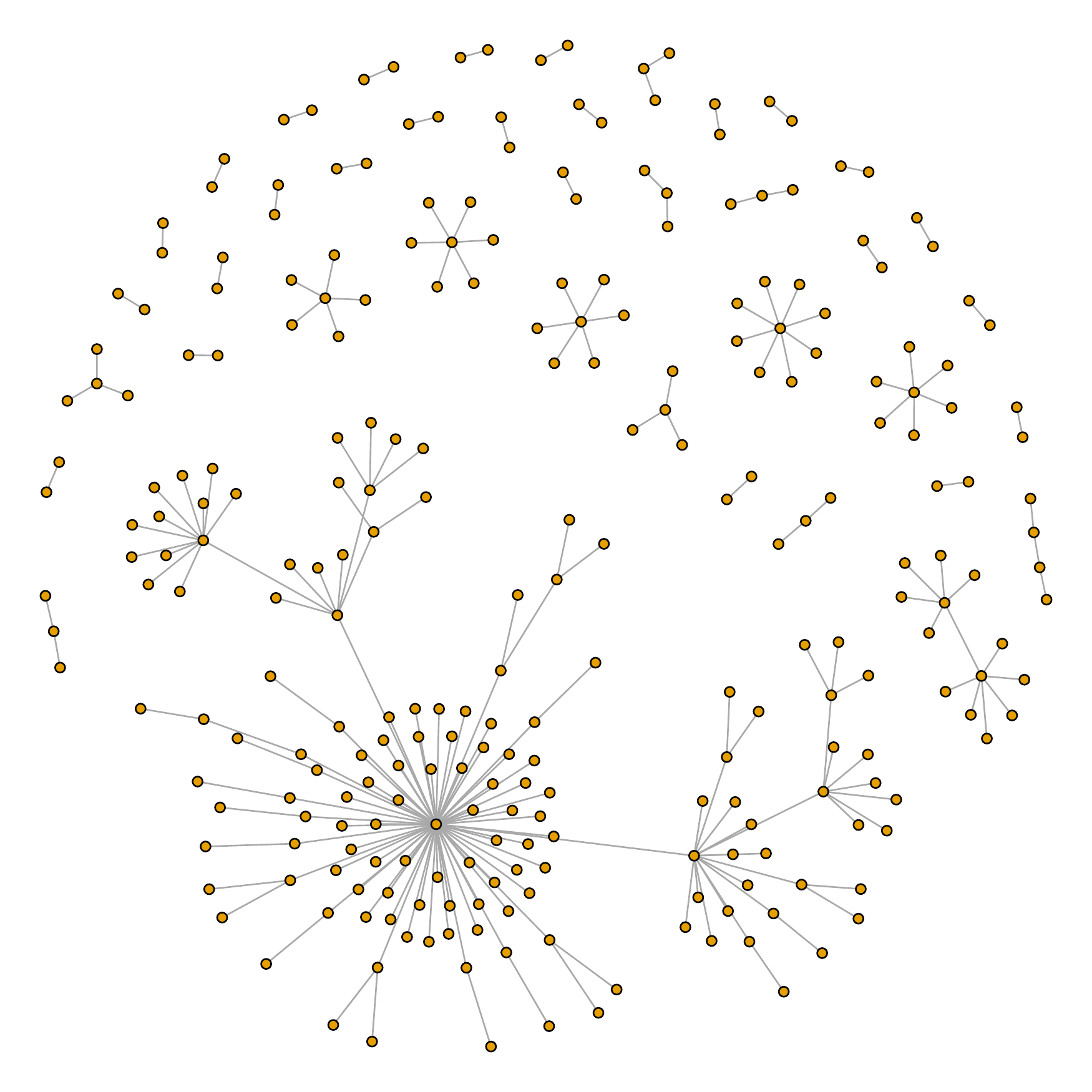}
    \caption{Topology} 
    \label{fig:mike_topology}
  \end{subfigure}
  ~
  \begin{subfigure}{0.3\textwidth}
    \begin{minipage}{\textwidth}
      \begin{center}
        \mbox{}\vspace{1.05cm}\mbox{}

Mike passed away!
        \mbox{}\vspace{.3cm}\mbox{}

Bye grandpa Mike
        \mbox{}\vspace{.3cm}\mbox{}

R.I.P.
        \mbox{}\vspace{.3cm}\mbox{}

How has television changed?
        \mbox{}\vspace{.3cm}\mbox{}

R.I.P.
        \mbox{}\vspace{.3cm}\mbox{}

\dots
        \mbox{}\vspace{.2cm}\mbox{}

      \end{center}
    \end{minipage}
    \caption{Text} 
    \label{fig:mike_text}
  \end{subfigure}
  ~
  \begin{subfigure}{0.3\textwidth}
    \includegraphics[width=\textwidth]{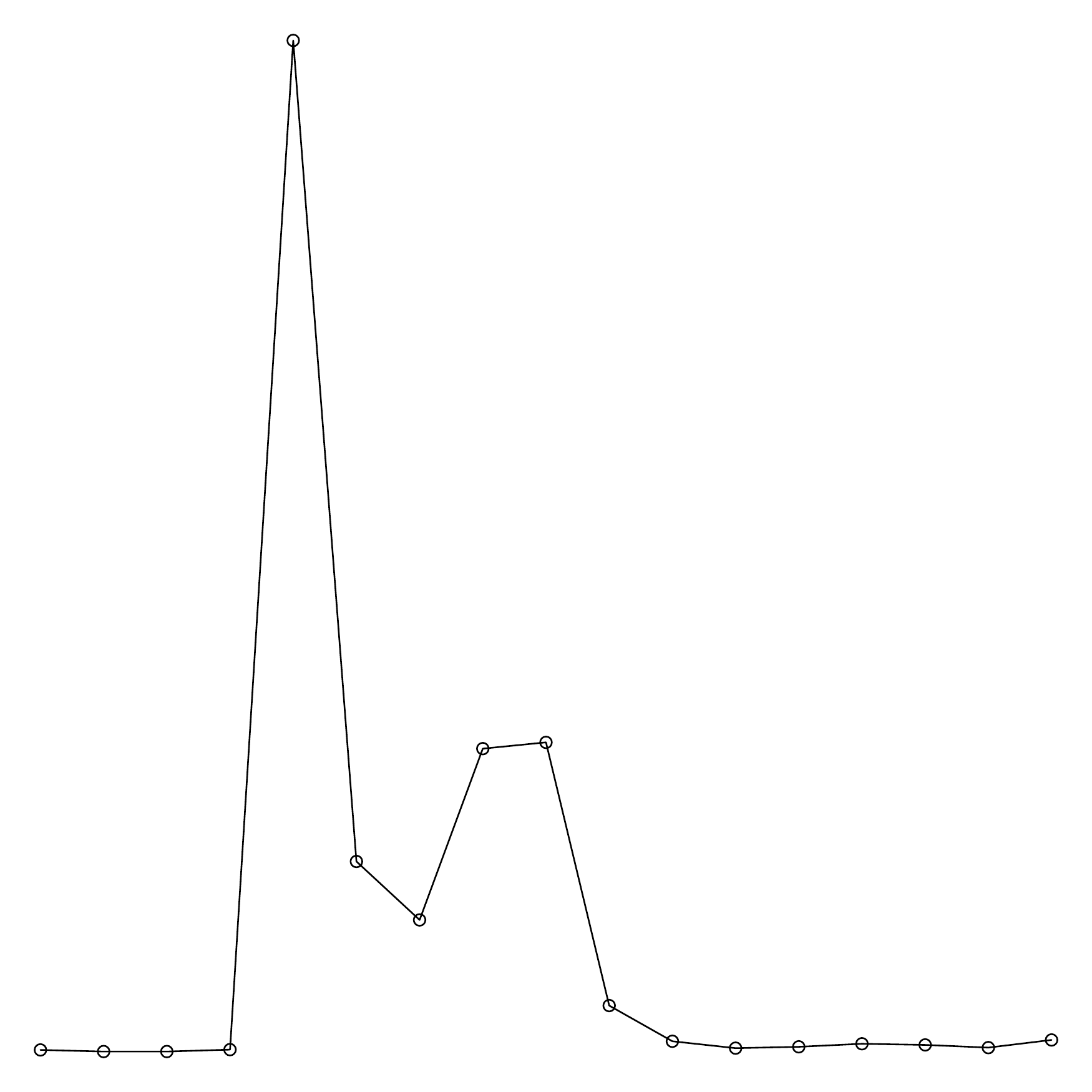}
    \caption{Time} 
    \label{fig:mike_time}
  \end{subfigure}
  
  \begin{subfigure}{\textwidth}
    \mbox{}\hfill%
    \includegraphics[width=.6\textwidth]{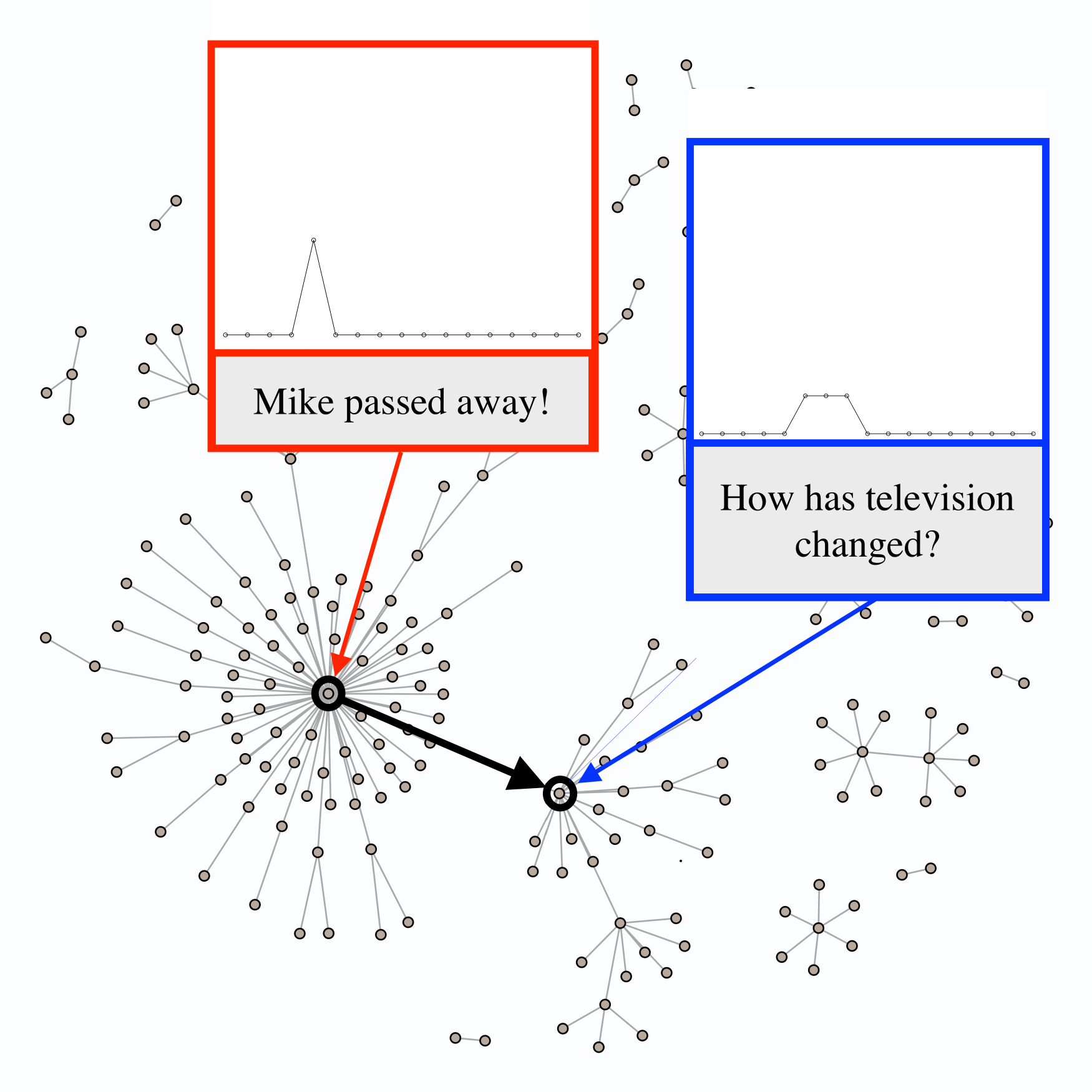}%
    \hfill{}\mbox{}%
    \caption{Topology, text and time} 
    \label{fig:mike_all}
  \end{subfigure}
  \caption{~\textbf{Three elements of an online human information network}:~\emph{a)} The topology, where each edge represents an observed information propagation path: user A writes a post about some news, user B reads the post and writes herself about it, for example by commenting on it;~\emph{b)} the text exchanged between users, that is, the text of posts and comments;~\emph{c)} the number of comments over time;~\textbf{d) Topology, time and text combined into a temporal text network}. Only details about two posts are shown.}
  \label{fig:mike}
\end{figure*}

One typical usage of social media data in research is to study how information propagates online. In one of the many studies on this topic, the authors have analyzed different aspects of the propagation process considering the online reactions generated by the death of a well-known Italian TV anchorman \cite{Magnani2010}.

In Figure~\ref{fig:mike} we have reproduced (a) the information propagation network, showing which posts contained information obtained by which others, (b) the text of some of the posts generated about this event, and (c) a temporal pattern indicating the number of comments per day.

While each of these pieces of information alone reveals something, putting them together into a temporal text network (Figure~\ref{fig:mike_all}) we obtain a much more comprehensive understanding of the process. \hlc{On the one hand, w}e can see that for the posts representing explicit attempts to propagate information (e.g., \textit{Mike passed away}) publication time is fundamental to determine their success, and only the first of this type of posts generated a large and sudden burst of reactions in a very short time; \hlc{on the other hand,} conversational posts evolving from it (e.g., \textit{How has television changed?}) can appear later and still create long but less dense chains of reactions. Other posts not present in the information propagation network neither explicitly give the news nor ask for an answer, generating no or few reactions, but still have the role of re-activating the information cascade so that even the latecomers can find a trace of it; some of these posts (e.g., \textit{Bye granpa Mike!}, or \textit{R.I.P.}) form what has been called an online mourning ritual.

In summary, time, text and topology together can lead to a deeper understanding of how this information network evolved into its current structure and how information propagated through it. 

\subsection{Contribution and outline}

In this work we introduce a simple but expressive and easily extensible model for temporal text networks, and define two main approaches to analyze this type of data. We also show how existing primitive data manipulation operations for multilayer networks can be composed to easily construct new algorithms for temporal text networks.

Our claim is that such a model can play a similar role of other recent attempts to unify related areas of network science, such as multilayer networks, which have boosted research in already existing fields (e.g., multiplex network analysis) by showing that results in one area could be directly applied to other types of data now expressed using a uniform terminology and mathematical form. \textbf{Our objective is to define an essential model, with a minimal number of features, so that several existing models can be unified into it without a significant increase in model complexity}.
We also believe that a unified model will promote the development of software libraries providing different data analysis functions for temporal text networks inside a single system, from centrality measures to community detection and generative models.












The article is organized as follows. In the next section we present an overview of related work, highlighting how a large \hlc{amount of research} has been produced to analyze human information networks\hlc{. As the main objective of this article is to introduce a data model for temporal text networks, our overview of the state of the art focuses on the data models already introduced in the literature, to allow a precise comparison with our model.} In Section~\ref{sec:model} we define our model as a simple attributed bipartite network. We also show how this simple model can be used to represent many existing types of text\hlc{-based} interactions, such as direct messages, multicast and broadcast. \hlc{In addition}, we show how to express \hlc{different types of information networks} using our model, and how to extend it with additional features. \hlc{Finally, we provide a detailed comparison of our model with the ones presented in the state of the art, showing how some existing models can be expressed using ours, while others can be obtained by applying some lossy processing to ours, e.g., replacing the exchanged text with a bag of words, a set of topics, a sentiment, etc. } Section~\ref{sec:analysis} explains how the model can be used in data analysis. We show how the \emph{direct} manipulation of the model can be complemented by two additional types of analysis: \emph{continuous} and \emph{discrete}. In the continuous case, time and text are treated as points in a metric space, and analysis operations are based on the computation of similarities between these points. In the discrete case, discretization operations (such as time slicing and topic modeling) are applied, encoding text and time into multiple discrete layers and enabling the direct application of the large number of methods already available for multilayer networks. In Section~\ref{sec:case} we present a practical example of our model and analysis strategies applied to Twitter data.

\section{Related work}
\label{sec:sota}


Our concept of temporal text network is a combination of text, network topology and time. 
\hlc{In the literature there is a large number of models supporting one or more of these aspects, and the objective of this section is to characterize existing models from a common viewpoint. In this way, in the next section we will be able to provide a precise comparison between our proposal and existing work, showing that our model is more expressive but at the same time consistent with existing approaches, reusing existing modeling constructs when possible. In particular, we will show that we can express existing models using ours, but not vice-versa.}

\hlc{Notice that t}here are \hlc{entire} well-established disciplines developed to address \hlc{text, network topology and time} in isolation, and we do not review these here as they are widely covered by text books \cite{Newman2010,Baeza-Yates:1999:MIR:553876}, described in numerous research papers (see for example \cite{Blei:2003:LDA:944919.944937} and subsequent extensions), and included in several software packages and systems. \hlc{Instead} we describe recent research efforts combining at least two of these aspects. 

\begin{figure*}[h!]
\begin{center}
\includegraphics[width=.8\textwidth]{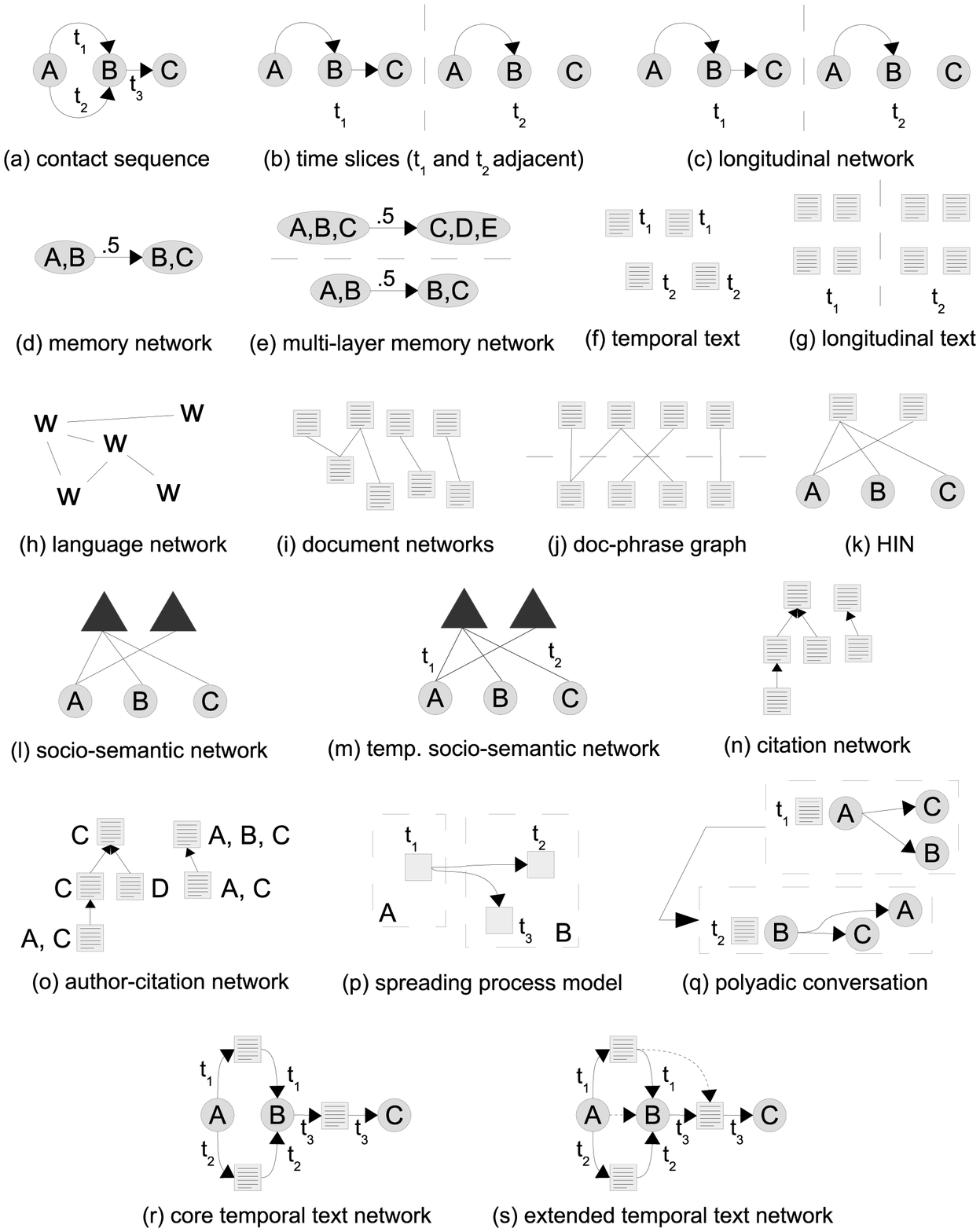}
  \caption{\textbf{\hlc{A visual gallery of models for time text and networks}.}}
  \label{fig:models}
\end{center}
\end{figure*}

\hlc{Table~\ref{tab:sota} presents a summary of the models selected for this review, also including our proposed model (core and extended temporal text network), organized according to four main criteria: (1) the type of graph used to represent the topological portion of the data, (2) the type(s) of nodes allowed in the graph, (3) the way in which text is represented in the model and (4) the way in which time is represented in the model. In Section~\ref{subsec:comparison} these criteria will be used for a comparison with our model. As our aim is to comprehensively list models, not papers, and the number of works using some of the models is very large, we have sometimes arbitrarily and unavoidably chosen a key set of references based on our knowledge and personal selection. Therefore, please notice that in the table we only indicate selected representative references; additional references are included in the text.
Figure~\ref{fig:models} complements Table~\ref{tab:sota} providing a visual intuition of the reviewed models and of the new models introduced in this article.}

\begin{table*}[!h]
  \centering
  \footnotesize
  \caption{\hlc{\textbf{Comparison of models} representing two or more of the main aspects of temporal text networks. The graph type is indicated as D: directed (undirected if D is not specified), O: ordered, G: Graph, MG: Multi graph, BG: Bipartite graph, ML: Multilayer graph. Node types indicate the domain of the nodes, and we distinguish between A (nodes used to represent actors) and X (nodes used to represent text-related objects). Given the variety of existing models, X is broadly used to represent full text documents, parts of it (phrases, words), other representations of documents such as bags of words (BoW), and also objects obtained by analyzing the text, such as concepts/topics. In this table we only indicate  selected references.}}
  \label{tab:sota}
  
  \begin{tabular}{llllll}
    \toprule
    
     {\bf Name } & {\bf Graph type} & {\bf Node types} & {\bf Text repr.} & {\bf Time repr.} & {\bf Refs.} \\
    

    \midrule
    
    Contact sequence & DMG & $A$ & --- & mostly edges &  \cite{Holme2012:Temporal,Gauvin2013}\\ 
    
    Time-slice & OML & $A$ & --- & layers & 
\cite{Mucha876} \\

    Longitudinal & OML & $A$ & --- & layers & 
\cite{snijders_2005,Snijders2014} \\

    
    
    Memory & DG & $A^n$ & --- & edges (implicit) & \cite{scholtes2014:temporal,rosvall2014:temporal,lambiotte2015:temporal,peixoto2017:temporal} \\
    
    Memory (multilayer) & DML & $A^1 \cup \dots \cup A^n$ & --- & edges (implicit) & \cite{scholtes2017:topology} \\
    
    

    
    
    
    
    Temporal text & --- & $X$ & document & vertices & \cite{BrucatoMontesi2014temporal} \\
    
    Longitudinal text & --- & $X$ & document & layers & \cite{OConnorBRS10ICWSMsentiment,Dodds2010} \\
    
    
    
    Language networks & G/DG & $X$ & word & --- & \cite{sole2010language} \\
    
    Document networks & G/DG & $X$ & document & --- & \cite{journals/jmlr/ChangB09,Menczer5261} \\
  
    Document-phrase graph & BG & $X \cup X$ & document, phrase & --- & \cite{Ren2017:CDA} \\
        
    HIN & BG & $A \cup X$ & BoW, concept, doc. & --- & \cite{Chang:2009:CLA:1557019.1557044,Wang2017:Models,Kralj2016:EnrichedHCIN} \\

    Socio-semantic network & BG & $A \cup X$ & concept & --- & \cite{Roth2017:FCA,Loet2017:SocioSemanticNetworkModel} \\
    
    
    Temp.~Socio-semantic network & BG & $A \cup X$ & concept & edges & \cite{Roth2010:SemanticCoevolution} \\
    
    Citation network & DG & $X$ & document & vertices & \cite{institute1964use,DBLP:journals/corr/cs-DL-0309023} \\
    
    Author-citation network & DG & $2^A \times X$ & document & vertices & \cite{white1981author} \\
    
    Spreading process & DG & $A \times X$ & --- & edge (delay) & \cite{Leskovec2007} \\
    
    Polyadic conversations & DG & $A \times 2^A \times X$ & document & vertices & \cite{Magnani2012:ConversationRetrieval}
    \\
    
    Core temporal text network & DBG & $A \cup X$ & document & edges & \\
    
    Ext.~temporal text network & ML & $A \cup X$ & document & edges & \\
    
    \bottomrule

  \end{tabular}
\end{table*}

\subsection{Time \& Topology}

\hlc{The most basic family of models including both time and topology is the \emph{contact sequence} \cite{Holme2012:Temporal,Gauvin2013}.}
\hlc{This is the most popular model for representing time and relations as a simple network structure. Mathematically, the model can be represented as a directed multi-graph $G = (V, E, T)$ with attributed edges. The set of vertices $V$ represent actors (e.g., individuals, companies) and the set of edges $E$ represent the interactions among the actors. When used in practice~\cite{Lambiotte2013burstiness,Gauvin2013,Paranjape2017:TemporalMotifs} the duration of the interactions is sometimes considered negligible and hence represented as a single scalar $t \in T$, while in other occasions the temporal information is represented as time intervals $t = (t_s, t_e)$ indicating when the contact between two actors starts ($t_s$) and ends ($t_e$) \cite{viard2016:temporal}. }
Contact sequences have been \hlc{typically} used to study information spreading~\cite{Lambiotte2013burstiness,Cheng2016:MethodsCascades}, \hlc{and existing concepts such as motifs and triadic closure have been re-defined to study the evolving structure of these networks}~\cite{Paranjape2017:TemporalMotifs,viard2016:temporal,Kim2017:TimeTriadicClousure}.


\hlc{
Differently from contact sequences, where interactions are time-annotated one by one, other types of models use sequences of time-annotated graphs, where each graph is  sometimes also called layer. In \emph{time-sliced} models, also known as time-aggregated models, time is expressed as an interval and an edge indicates that an interaction has happened at some point during the time interval associated to the graph \cite{Mucha876}. These models are typically obtained starting from a contact sequence and aggregating edges by time. In}
\emph{longitudinal networks} relationships about the same or similar actors are detected at different \hlc{points in} time  \cite{snijders_2005,Snijders2014}. \hlc{From a data modeling point of view, time slicing and longitudinal networks are very similar, and in practice the main difference lies in the nature of the time annotation associated to each slice, where in time slicing adjacent slices are typically associated with adjacent time intervals while in longitudinal network studies adjacent layers represent network snapshots obtained at specific points in time. Different types of time annotations are described for example in \cite{batagelj2016:temporal}.}

\hlc{Memory models provide a different view over a temporal network, where ordered tuples of two or more actors are represented as single nodes \cite{scholtes2014:temporal,rosvall2014:temporal,lambiotte2015:temporal,peixoto2017:temporal}. For example, second order memory networks~\cite{scholtes2014:temporal,rosvall2014:temporal} can model the impact of one predecessor edge. 
For example, if an actor $v_i$ is receiving one message from $v_j$ and one from $v_k$, and is later sending a message to $v_j$ and one to $v_k$, a contact sequence loses information on whether $v_i$ is replying to $v_j$ and $v_k$ ($j$ $\rightarrow$ $i$ $\rightarrow$ $j$, $k$ $\rightarrow$ $i$ $\rightarrow$ $k$) or forwarding the messages ($j$ $\rightarrow$ $i$ $\rightarrow$ $k$, $k$ $\rightarrow$ $i$ $\rightarrow$ $j$). A first-order memory model will contain nodes for each pair of users and have an edge between two nodes if the corresponding pairs appear on consecutive paths. In our example, if $v_i$ is replying we will have two edges in the memory model: ($\overrightarrow{ji}$, $\overrightarrow{ij}$) and ($\overrightarrow{ki}$, $\overrightarrow{ik}$), while if $v_i$ is forwarding the messages we will have the edges ($\overrightarrow{ji}$, $\overrightarrow{ik}$) and ($\overrightarrow{ki}$, $\overrightarrow{ij}$).
Higher order memory networks also exist~\cite{lambiotte2015:temporal}, although they are not as common, to represent causality effects between pathways consisting of 3 or more nodes. Deciding the order of the model is not trivial as specific patterns can be revealed only on a specific subset of memory models. To solve this problem, Scholtes et al. \cite{scholtes2017:topology} introduced a multilayer memory network, composed of multiple memory networks of different order hierarchically connected between them (e.g., each node in the 2nd-order layer $v_{ij}$ is connected with all nodes in the 3rd-order layer whose path $v_klm$ contains the leg $\overrightarrow{ij}$, so $\overrightarrow{ij} \subseteq \overrightarrow{klm}$).}


\hlc{Time often plays an important role when networks are concerned, because networks often represent dynamical systems. However, in Table~\ref{tab:sota} we have only listed distinct data models explicitly providing time annotations. As an example,
}
\emph{growing network models} \cite{Newman2010} such as preferential attachment \cite{Barabasi99emergenceScaling} aim at explaining the observed topology of empirical networks based on how they evolve in time from an initial small network. \hlc{Even if nodes and edges join the network one after the other, there is no explicit representation of time in the final model. Similarly, we have not listed papers about methods not explicitly introducing new data models, such as~\cite{lentz2013:temporal}.}

\subsection{Time \& Text}

Time is often present inside text, and commercial systems handling large human information networks from Google mail to common text messaging applications on smart phones can automatically identify \hlc{the messages} and annotate the text with temporal information.

In research, text and time are studied together in the field known as temporal information retrieval \cite{Alonso:2007:VTI:1328964.1328968,Kanhabua:2015:TIR:2864701.2864702}. This is an active area, also represented at the TREC conference where state-of-the-art information retrieval methods compete on various practical tasks. Time can be present in the text, as in the examples above or as metadata, expressed as absolute or relative time and it can also be specified in queries used to express information requirements \cite{BrucatoMontesi2014temporal}.

Another set of studies has focused on how text evolves in time, and in particular sentiment, with case studies ranging from tweets \cite{OConnorBRS10ICWSMsentiment} to songs, blogs and presidential speeches \cite{Dodds2010}. \hlc{Text and time are also studied across data sources,} for example to correlate texts from online news to trends emerging in time series such as financial data \cite{Lavrenko00miningof}. \hlc{However, no specific data model is used for this type of tasks, but only time-annotated documents (understood in a broad sense, including words, etc.) and time series.}

\subsection{Text \& Topology}

Text and networks have been studied together in various areas, either without considering time or using networks to represent relationships between texts.

Models where nodes represent parts of a document have been used in structured information retrieval, which was a particularly active research area when hypertexts and markup languages became popular 
\cite{Kotsakis:2002:SIR:508791.508919}. Text is often contained inside some structure \hlc{(}e.g., a title, sections, sub-sections, etc.\hlc{)} and queries can be tuned to return specific parts of a document instead of a full one. As an example, if the searched keyword is contained inside Subsections 3.1 and 3.3 of a document, a query may return either the two subsections, or the whole Section 3, depending on the method.

\hlc{More relevant for this article are document networks, that are graphs whose nodes represent text documents \cite{journals/jmlr/ChangB09,Menczer5261}. These network models can be classified in different groups depending on whether they include time or not; later in this section we refer to citation networks as a type of directed document network where time is also typically present.} Text mining, and in particular clustering, \hlc{can be applied to document networks to identify groups of documents that are similar not only because of their text but also because of their connections, as summarized in a} recent article about clustering attributed graphs \cite{Bothorel2015}.

Several works have focused on networks extracted from text, \hlc{and we can broadly classify them into models representing the text itself, aimed at characterizing language, and models representing actors and concepts mentioned in the text.}

Networks where nodes represent words have been used to model both text documents and languages \cite{sole2010language}. For example, a document can be modeled as a network where words are connected by an edge when they are contiguous, or appear in the same sentence, paragraph, etc. Similarly whole languages can be modeled focusing on the relationships between words, as in WordNet or BabelNet.

\hlc{With regard to the second class of models for networks extracted from text,} Named Entity Recognition methods \hlc{are typically used} to identify the nodes and co-occurrence (or other language analysis approaches) to create edges among them \cite{Diesner2004RevealingSS,Chang:2009:CLA:1557019.1557044}. In this case, the output network connects different portions of a text document, or concepts extracted from the text.

A model that has been used to represent the relationships extracted from texts is \hlc{known as} heterogeneous information network (HIN)~\cite{Shi2017:HCISurvey,Ren:2016:AER:2872518.2891065}. \hlc{HINs are defined as attributed directed graphs $G = (V, E, A, R)$ with an object type mapping function $V \rightarrow A$ and a link type mapping function $E \rightarrow R$, so that each object in the network (vertices and edges) belongs to a single type and if two edges belong to the same relation type $R$, the two edges share the same starting object type as well as the ending object type. For example, HINs have been used in the past to model co-occurrence relations between entities (e.g., famous characters, sports, companies) in Wikipedia articles~\cite{Kralj2016:EnrichedHCIN}. In~\cite{Chang:2009:CLA:1557019.1557044} vertices represent either famous characters from the text or bags of words, while the edges connect words that best explain the contexts where two or more famous characters appear together in the text. Document-phrase graphs as defined in \cite{Ren2017:CDA} are also HIN-based models, and more in detail probabilistic bipartite networks $B = (V, U, E, W)$ where the vertices in one partition $V$ represent documents from a large document collection, the vertices in $U$ represent salient phrases which are semantically relevant to one or more documents in $V$, and edges $E$ indicate the relevance of each sentence for each document. }
\hlc{
HINs are not limited to represent relations within documents, text and concepts; but they can also model relations between actors and text. The most common use of HIN is actually to represent co-author or citation networks. In~\cite{Wang2017:Models}, for example, the authors use an heterogeneous information network to describe the relations between scientific articles, their authors, and the venues where they were published.} 



One of the concerns recently raised against using methods from social network analysis to analyze social media is their intrinsic actor-centered approach (e.g., people, companies, stakeholders), focusing  on social interactions without properly characterizing other aspects of the communication~\cite{Roth2017:FCA}. 
A similar argument can be used against the use of just Natural Language Processing or semantic networks~\cite{sowa2014principles}.

Following this reasoning, a recent stream of research focused on combining structural and semantic data simultaneously,
which led to the formalization of the socio-semantic network model~\cite{Roth2010:SemanticCoevolution,Roth2017:FCA,Loet2017:SocioSemanticNetworkModel}. Originally, socio-semantic networks were just bipartite graphs interconnecting \emph{agents} (also known as actors in Social Network Analysis) with semantic objects called \emph{concepts}, corresponding for example to terms, n-grams, or lexical tags.

During the last decade the socio-semantic network model has been extended to extract more valuable knowledge from social media. An illustrative example of such extension can be found in~\cite{Loet2017:SocioSemanticNetworkModel} where the authors propose to combine the aforementioned social and socio-semantic networks into a single model. In short, they use a single matrix representation where the diagonal sub-matrices represent the relation between the same type of entities (agents and concepts) and the off-diagonal matrices represent the relation between different ones (agent/concept and concept/agent). \hlc{From the point of view of data modeling, HINs are very related to socio-semantic network models, even though HINs have been introduced as more general modeling tools while socio-semantic networks have emerged and are used in a specific application context.}

A final work worth mentioning in this class is \cite{Rosen-Zvi:2004:AMA:1036843.1036902}, where topic modeling is performed using an extended model considering not only the association between topics, words and documents, but also the association between documents and their authors. \hlc{However, this has not been included in our summary table because it introduces a generative model to summarize the data in the form of parameters indicating the probability that a given actor produces a given set of words, but not to represent the empirical data showing which actors have written what text.}

\subsection{Time \& text \& topology}

Many works in the literature have dealt with time, text and topology using ad hoc models specifically designed to capture relevant aspects of specific platforms such as Twitter. 
For example, in~\cite{Tamine2016:IntroInterestText} a communication network is built in three steps: (1) conversation trees are extracted from the dataset by inversely following the chain of Twitter user interactions (replies, mentions and retweets); (2) the trees are pruned based on the time elapsed between the root \hlc{t}weet and the overlap of tweets and participants in the tree; (3) finally, all trees are merged to generate a simple weighted graph of interactions between authors. 
\hlc{A related model is the so-called polyadic conversation~\cite{Magnani2012:ConversationRetrieval}, designed to describe user interactions in microblogging sites as a series of related conversations --- also called polyadic interactions. A polyadic interaction is a tuple $i = (v, U, m, t)$ where $v \in V$ is the sender of the message $m \in M$, $U \subseteq V$ is the set of receivers and $t \in T$ is the timestamp of the communication act. A polyadic conversation is then defined as a chronologically ordered tree $G = (I, E)$ where $I$ is a set of polyadic interactions and $E \subset I \times I$. }

In~\cite{Roth2010:SemanticCoevolution} a temporal model was used to compare the co-growth of two epistemic networks, a Twitter dataset and a set of related blogs, with the underlying social network of contacts. The temporal information attached to the edges of the network is, afterwards, used to compare the order of formation of epistemic and social communities. 

\hlc{Citation networks have received a lot of attention, and include text documents, directed edges between them and also time annotations
\cite{institute1964use,DBLP:journals/corr/cs-DL-0309023}. In addition, when author co-citation analysis is performed \cite{white1981author}, the underlying data model must also contain information about who authored which documents.}

\hlc{
Information diffusion processes are often modeled including the diffused information item (meme, blog post, etc.), the actors propagating it, and the times of propagation. This is for example the case for the model used in \cite{Leskovec2007}. However, the majority of these models do not use text to perform the data analysis, but (sometimes) to define the links between documents. Time can also be used to infer network structure based on the observation of propagation events. For example, the observation of a group of individuals repeatedly re-sharing common tweets in the same temporal order may suggest that these people are connected, and that information (tweets, in this case) passes through these hidden connections \cite{GomezRodriguez:2010:IND:1835804.1835933}. In \cite{Salehi2015survey} existing theoretical diffusion models for interconnected networks are reviewed, extending concepts in information diffusion to a multilayer model.}

In order to preserve as much original information as possible, \v{S}\'{c}epanovi\'{c} et al.~\cite{Goncalves2017:ModelsMethods} use a more generic process to build the network, mixing techniques from social network and semantic analysis. In their work, the communication network \hlc{is} modeled as a simple, temporal graph using the Twitter ``replies'' to relate actors with each other. Then, they appl\hlc{y} several semantic analysis procedures to generate supporting networks that describe the text\hlc{-}related features. A comparative analysis between the communication network and a subset of the semantic networks \hlc{is} used to study several aspects of the overall system such as semantic homophily and its evolution. \hlc{However, from a modeling point of view text is not explicitly represented in this model, but coded inside the semantic layers. We will later use a related approach to exemplify how to use our model for data analysis.}

Some attention has also been devoted to models describing co-evolutionary networks \cite{GrossBlasius2008coevolutionary,Magnani2013}. Some of these models allow the representation of a status associated to each node. Statuses can be used for example to represent the political affiliation of the person represented by the node. In growing network models, the status can influence the evolution of the network for example by increasing the probability that people will create connections with other individuals sharing the same political affiliation \cite{Kimura2008coevolutionary,Lee2005affinities}. \hlc{As for the case of simple network growing models, time is not typically kept at the end of the growing process, and in addition status has not been used to model text to the best of our knowledge. Therefore, we have not included these works in our summary table, even if we consider them potentially relevant for this field if extended in the future.}
 

\section{Modeling temporal text networks}
\label{sec:model}




In our opinion, a good model for temporal text networks should be general enough to be able to represent a wide range of systems, but also contain a minimal number of modeling constructs, to make the model easier to use and study. In other terms, a good compromise should be found between expressiveness and simplicity.
In addition, given the large number of existing models that have been used for a long time to describe specific aspects of temporal text networks, we believe that both the modeling constructs and the terminology used in our model should be as aligned with previous work as possible.
Following these design principles, we propose the following definition of temporal text networks:

\begin{definition}[Temporal text network]
\label{def:txt}

A temporal text network is a triple $(G, x, t)$ where:

\begin{enumerate}
  \item $G = (A, M, E)$ is a directed bipartite graph, where, $A$ is a set of actors, $M$ is a set of messages, and $E \subseteq (A \times M) \cup (M \times A)$.
  \item $x: M \rightarrow X$, where $X$ is a set of sequences of characters (texts).
  \item $t: E \rightarrow T$, where $T$ is an ordered set of time annotations.
\end{enumerate}

and where the following constraints are satisfied:

\begin{enumerate}
  \item $\forall m \in M, \textrm{in-degree}(m)=1$.
  \item $(a_i,m), (m,a_j) \in E \Rightarrow t(a_i,m) \leq t(m,a_j)$.
\end{enumerate}

\end{definition}

In our model edge directionality indicates the flow of text in the network: $(a_i, m_j) \in E$ indicates that actor $a_i$ has produced text $m_j$, while $(m_j, a_i) \in E$ indicates that actor $a_i$ is the recipient of text $m_j$.
Actors with out-degree larger than \hlc{0}  are information producers, actors with in-degree greater than \hlc{0} are information consumers, and actors with both positive in- and out-degree are information prosumers.
 
Text is represented as a combination of a text container ($m \in M$), and a textual content ($x(m)$). As a consequence, actors in our model do not only generate text, but produce text messages. Two text messages (for example, two tweets, or two emails) may be different messages even if they contain the same text and have been exchanged between the same actors at the same timestamp.


The third key component of temporal text networks is the time attribute $t$. In our model, time is defined based on a generic set of ordered time annotations $T$. This enables the adoption of several ways of representing time: as an absolute date-time, as a relative date-time, as a timestamp with an arbitrary format or as a discrete time interval if time has been sliced into time windows as it often happens when temporal networks are analyzed \hlc{(See Table~\ref{tab:sota})}.

When writing about the model's elements, we will sometimes use a concise notation. For example, we will sometimes write an edge and its time together, as in: $(a_i, m_j, t_q)$, where $t_q = t(a_i, m_j)$, and we will sometimes write a message by also indicating its sender, its recipients and its text, as in: $(a_s, m_j, \{a_{r_1}, \dots, a_{r_n}\},\textrm{``text''})$, where $\textrm{``text''} = x(m_j)$. Finally, when all the timestamps on the edges adjacent to a message are equal, we can also add a time to the previous notation, as in: $(a_s, m_j, \{a_{r_1}, \dots, a_{r_n}\},\textrm{``text''},t_q)$.

\subsection{Applicability}
\label{subsec:models}

While very simple, the model introduced above can be used to represent a range of different forms of communication and data from different sources. \hlc{In particular, by explicitly dividing the network nodes into \emph{actors} and \emph{messages}, their relations implicitly carry more information. For example, w}hether the type of communication implemented by a message is unicast, multicast or broadcast is 
indicated by the out-degree of the message.

With \textbf{unicast} 
a message such as a handwritten letter is sent from a single source to a specific target. This form of written communication has been preserved to the present day through 
instant messaging services such as those offered by Twitter, Facebook Messenger or Whatsapp \hlc{and, more traditionally, using the electronic email}. 
Unicast communication allows to keep some text private between two actors, but it can have a large overhead if the same text must be sent to multiple sources because it requires an individual message for every recipient. In order to reach a larger population it is sometimes preferable to use \textbf{broadcasting} or \textbf{multicasting}. In the former, the message is transmitted to all possible receivers\footnote{For simplicity we use the expression ``all  possible receivers'' to refer to the community in which the information is spread, independent of whether the community is the whole Internet, the whole world or a set of members registered to a private site.}, while
when the information is addressed to a group of people but not to all possible receivers, such as a post on a Facebook wall, the communication is called multicast.

\begin{figure}[h!]
\centering
  \includegraphics[width=0.45\textwidth]{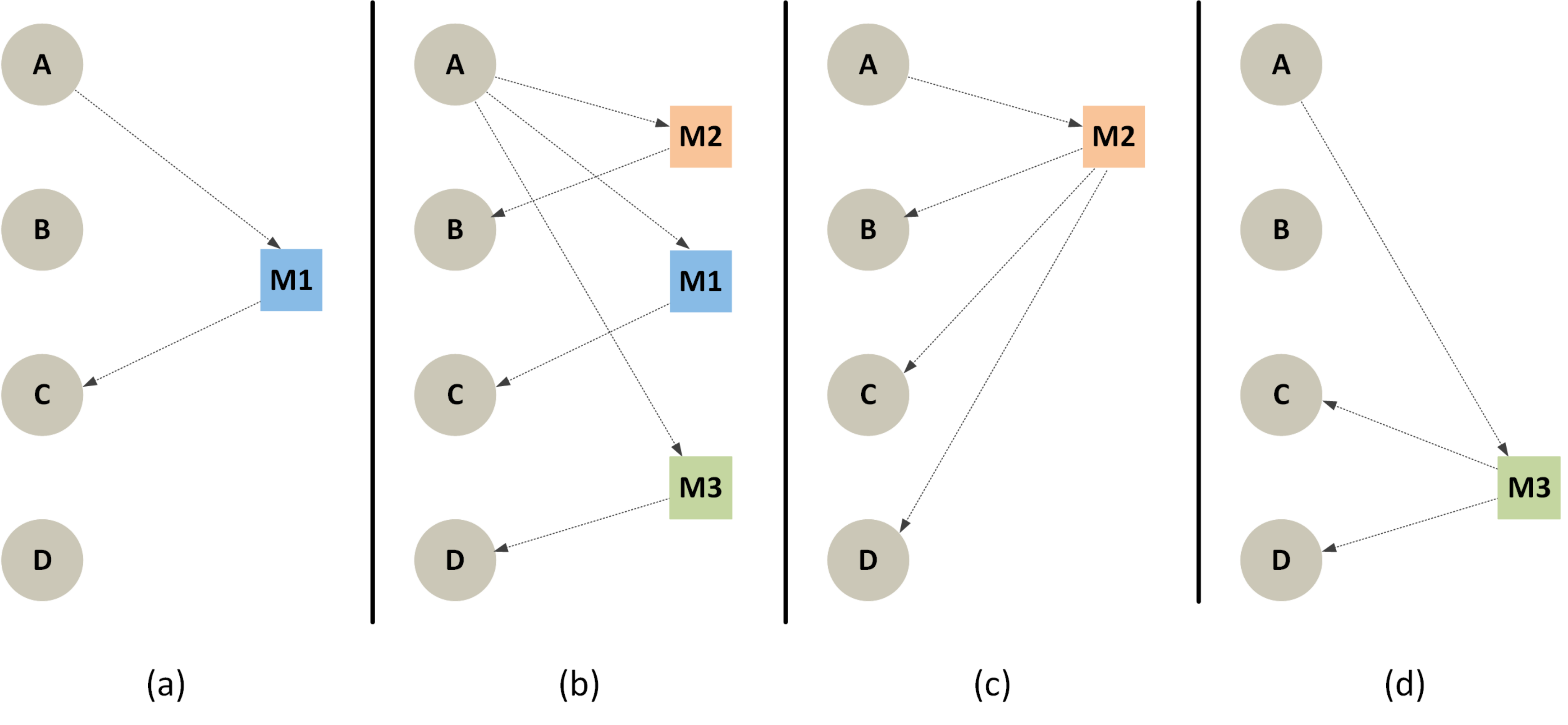}%
\caption{~\textbf{Models for different types of communication}.~\textit{a)} unicast from A to C;~\textit{b)} unicast from A to B, C and D;~\textit{c)} broadcast from A --- which can also be implemented as in the previous case if $x(M_1)=x(M_2)=x(M_3)$ and~\textit{c)} multicast from A to C and D.}
  \label{fig:com}
\end{figure}

Fig.~\ref{fig:com} shows these different types of communication represented using our model.

\begin{figure*}[th!]
\centering
  \includegraphics[width=0.95\textwidth]{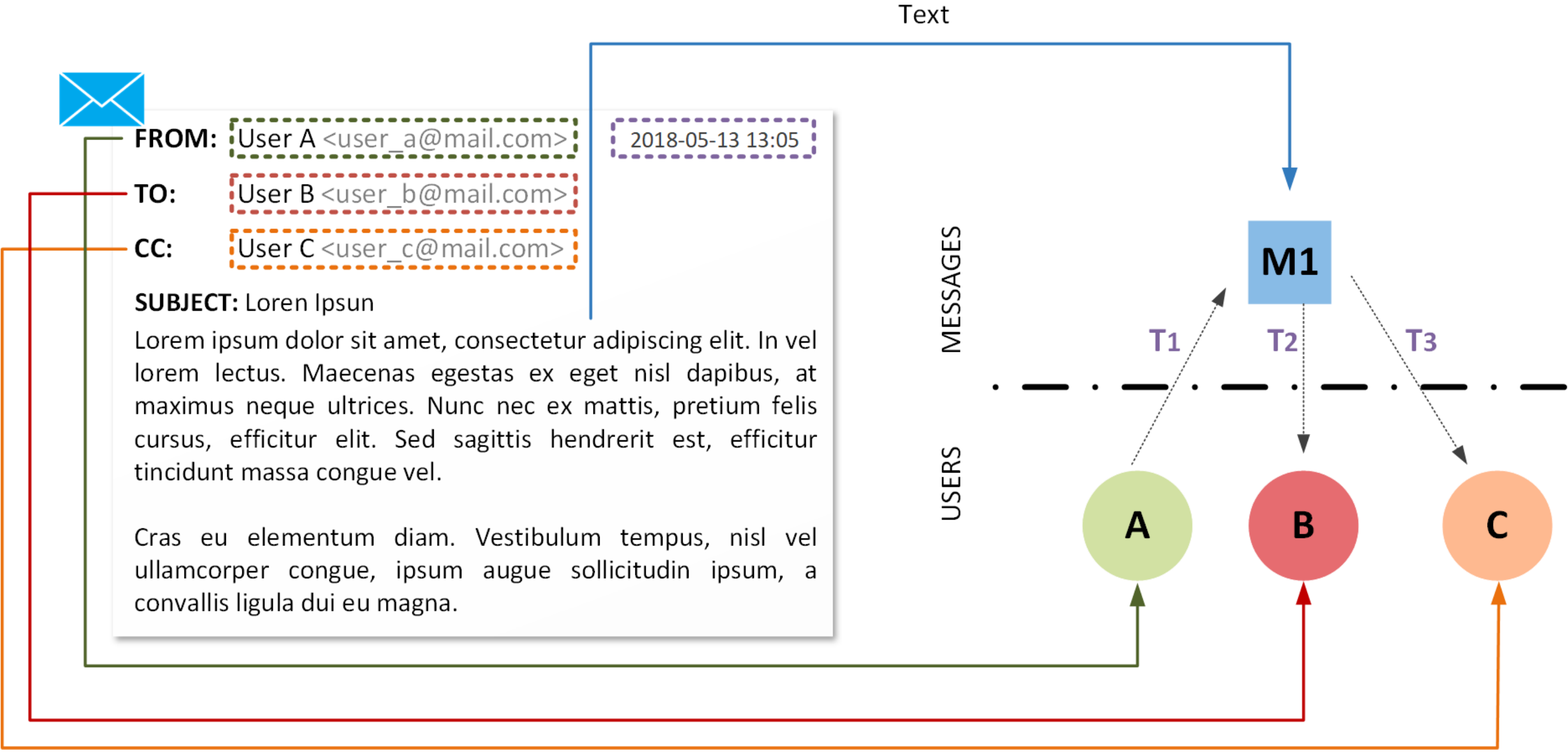}%
  \caption{\hlc{\textbf{Model of a multicast email as a temporal text network}. The entire text content of the email (including the subject line and the body) are encoded as a single message $M_1$. The sender of the email~\textit{(User A)} and the two friends (\textit{User B}, ~\textit{User C}) are modeled as individual actors. In this case, the ingoing and outgoing edges of the message contain a different time, indicating the delivery and reception timestamps registered in the email servers.}}
  \label{fig:mail_ttn}
\end{figure*}


Figure~\ref{fig:mail_ttn} shows an example of how \hlc{a multicast communication through email} 
can be modeled as part of a temporal text network. The resulting network includes the sender of the message (\emph{User A}) and two other actors (\emph{\hlc{User B}} and \emph{\hlc{User C}}) who where \hlc{explicit recipients} 
of the message. The fourth vertex $M_1 \in M$ represents the \hlc{email} 
and $x(M_1)$ corresponds to its text content \hlc{(the subject line and the body content).} 
In this case, \hlc{the time attribute associated to each one of the edges represents the time when the message was delivered or received by the SMTP and POP3 servers allowing us not only to represent the communication flow, but also the effect of the channel and/or medium.}
Representing multiple \hlc{emails} 
as in the example above would lead to a full temporal text network. 

\hlc{In the next section we describe how to express other human information networks by extending our core model.}


\subsection{Model extensions}
\label{subsec:extension}

\begin{figure*}[th!]
\centering
  \includegraphics[width=0.95\textwidth]{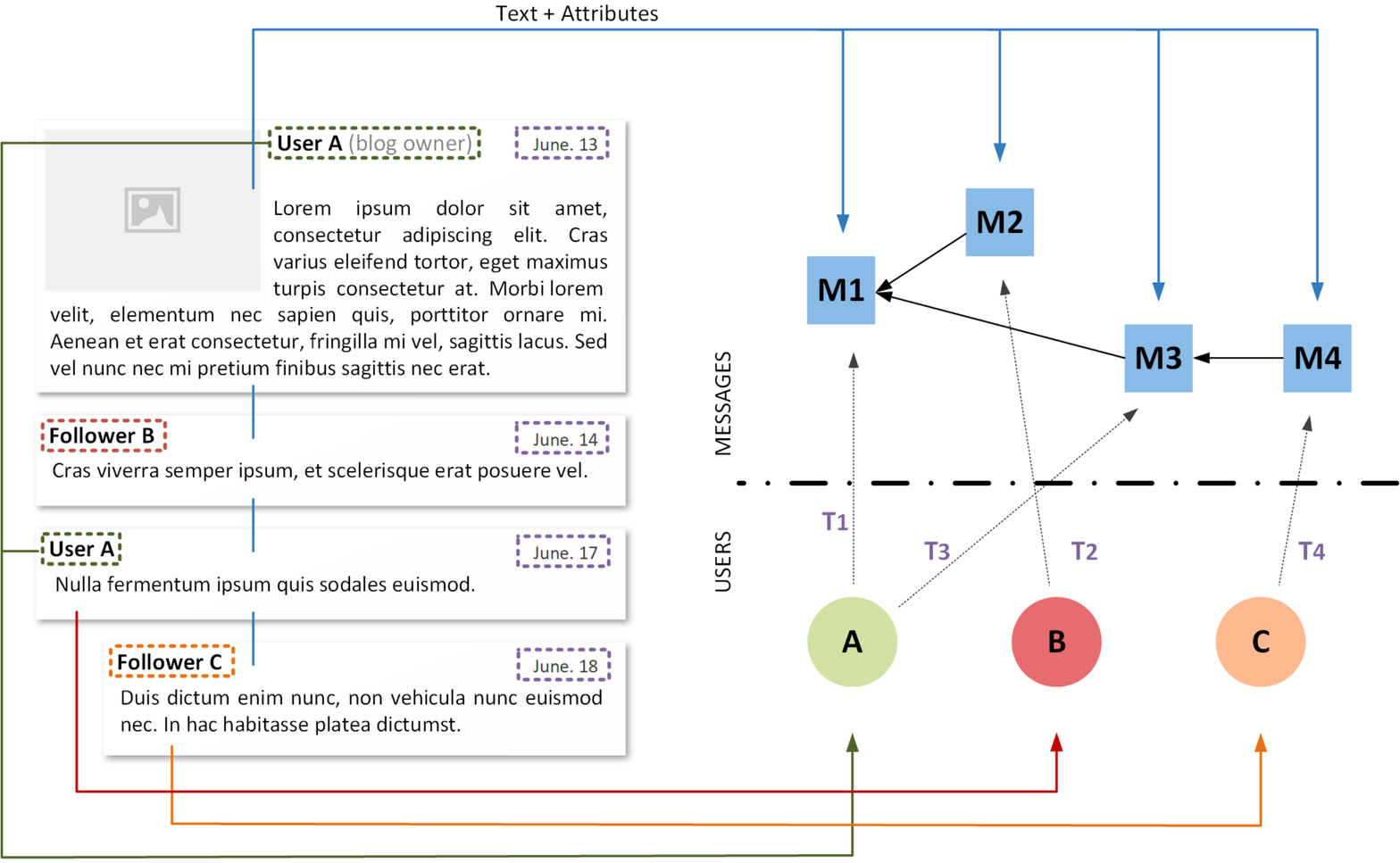}%
  \caption{\hlc{\textbf{Model of a blog post as a temporal text network}. The original data set contains a blog post $M_1$ and three comments ($M_2, M_3, M_4$); which are encoded as three individual messages. The three participants on the discussion (\textit{User A}, ~\textit{Follower B}, ~\textit{Follower C}) are modeled as individual producers. In this case, the edges of the messages indicate the relation between their content.}}
  \label{fig:blog_ttn}
\end{figure*}

\begin{figure*}[th!]
\centering
  \includegraphics[width=0.95\textwidth]{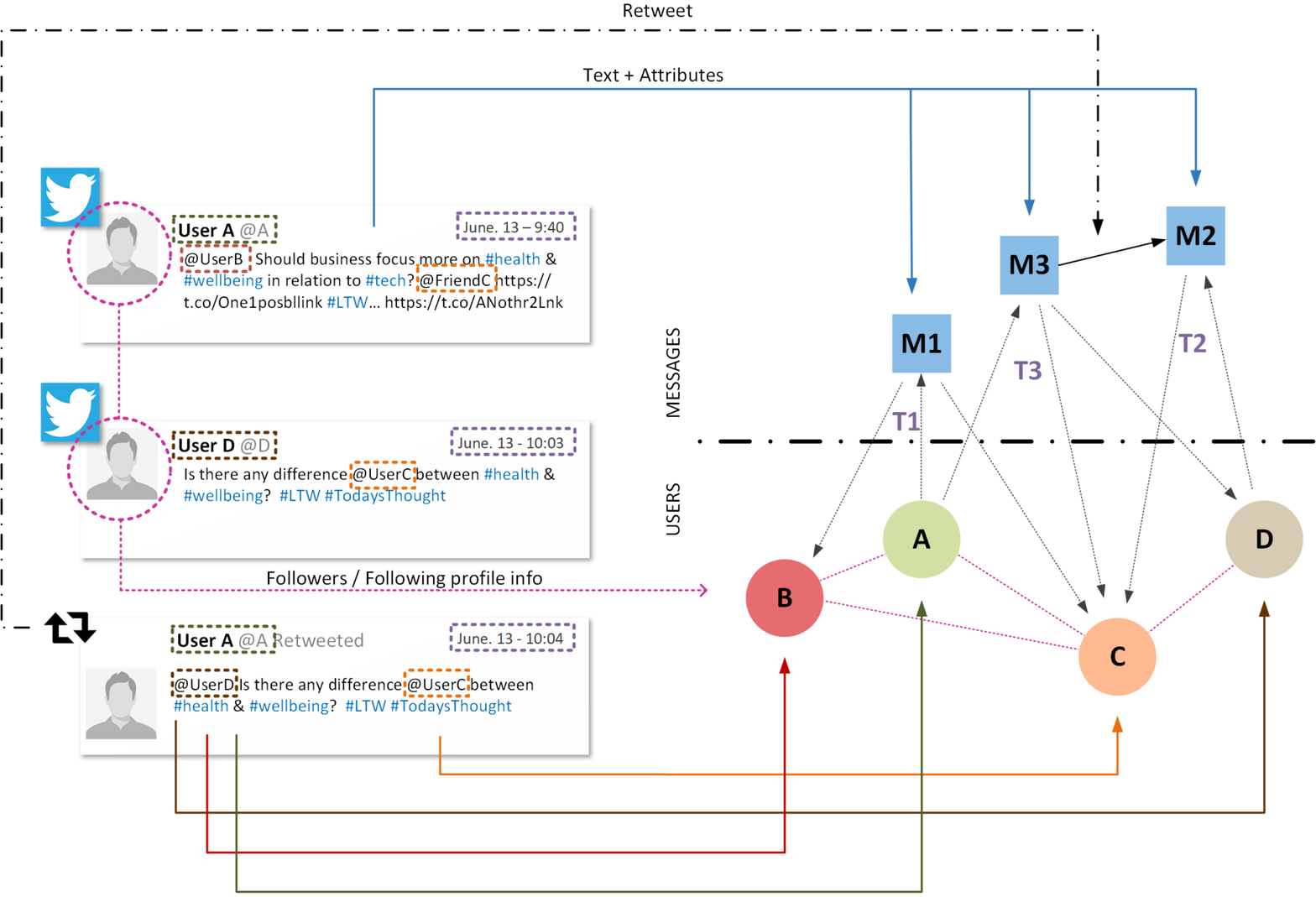}%
  \caption{\hlc{\textbf{Model of a Twitter network as a temporal text network}. The entire content of each tweet (including hashtags, urls and retweeted content) are encoded as messages. Senders (\textit{@A}, ~\textit{@D}) and mentioned users (\textit{@B}, ~\textit{@C}, ~\textit{@D}) are modeled as individual actors. In this case, both the ingoing and outgoing edges of the message contain the same time, which indicates when the tweet has been sent. The edge between $M_3$ and $M_2$ indicates the retweet relation between both tweets.}}
  \label{fig:twitter_ttn}
\end{figure*}

One of the design principles we used to define our model was simplicity, to make it tractable and general. On top of the basic model defined above, we can also easily add extensions to fit context-specific requirements.


With regard to the structure, we can straightforwardly add edges between \hlc{messages} to represent either information available from the data such as retweets on Twitter, or information deduced from the analysis of the data such as links indicating that one message is probably an answer to another, if we want to study information flows. \hlc{Figure~\ref{fig:blog_ttn} shows, for example, the modeling process of a blog post $M_1$ and the associated comments from the readers $\{M_2, M_3, M_4\}$. In this particular case, we know the identity of each one of the authors, because they are authenticated in the web platform, but we do not know exactly who are the recipients of their comments. While we can assume by context that the blog post $M_1$ was read by follower $B$ and that her message was then read by the blog owner $A$, it is uncertain what the third user (follower $C$) has read. We only know that the text produced by user $C$ is a reply to the previous comment $M_3$, but we cannot infer if he has or has not read the previous messages $M_1$ and $M_2$. One possible way to model such scenario is to represent the relation between messages instead of the relation between messages \emph{and} receivers. Similarly, in the example of Figure~\ref{fig:twitter_ttn} the edges between messages are used to represent retweets on a micro-blogging platform.}

As we discuss in the next sections, this type of extension would nicely fit our analysis framework where one main class of operations transforms the data into a multilayer representation. Similarly, we may add edges between 
\hlc{actors indicating other types of relations relevant for the analysis of the human information network such as indirect recipients. Figure~\ref{fig:twitter_ttn} shows the modeling process of Twitter as a temporal text network. Unlike the previous communication channels we discussed, in Twitter the recipients of the information are encoded in the text of the messages rather than being explicit in the metadata (e.g., the edge ($M_1, B, t_1$) exists because actor $A$ mentions $B$ in the first message of the data set). In addition, Twitter users can also see messages from other users they are following, which in our model is represented by the actor-to-actor relations. This difference between intra- and inter-layer relations allows us to differentiate between direct and indirect communication in many social platforms.}

In our basic model $x$ represents a generic string of characters over some alphabet, whose interpretation will depend on the source of the data and the context of the analysis. For example, while the symbol \emph{\#} usually denotes the start of a filtering tag in online social networks such as Twitter or Instagram, in  other media sites it is just an acronym for the word ``number''. Therefore, for specific application contexts additional attributes can be added for example to messages by providing special information, such as the hashtags included in the text in the case of Twitter (See Figure~\ref{fig:twitter_ttn}). In particular, we can think of having three types of information associated to each message:

\begin{enumerate}
  \item The text, as in our basic model,
  \item Metadata that is available in the specific data source used for the analysis, such as links to other resources (webpages, other tweets or multimedia content), like and retweet counts, or hashtags. 
  \item Additional information not directly available from the data source but obtained analyzing the text, for example through topic analysis. 
\end{enumerate}


Different types of temporal information have been used in existing works on temporal networks and temporal text analysis (See Section~\ref{sec:sota}). For example, time can represent actions from the users such as the time when a message is posted and\hlc{/or} the time when it is read \hlc{as we did in the Twitter example}. Alternatively, \hlc{times can be used to represent a physical property of the channel, as it happens in computer networks when there can be a transmission delay from the source to the destination of a message (See Figure~\ref{fig:mail_ttn})}. Finally, time can also be associated to the message, indicating for example the time interval when the message exists. Furthermore, this information can be complete or incomplete, so that if only the initial time of the interval exists we must assume the message is still valid at the time of analysis \hlc{as we did when we describe the blog posts}; it can be private (accessible only to specific actors) or universally accessible by everyone.

\hlc{

\subsection{A comparison with the state of the art}
\label{subsec:comparison}

Our core and extended models of temporal text networks allow us to describe a variety of human information networks ranging from person-to-person email communication to complex interactions in social media sites. In Section~\ref{sec:sota} we summarized other models from the literature, that have been used in the past to partially support similar scenarios. In this section we provide a comparative review between our models and the ones described in Table~\ref{tab:sota} and Figure~\ref{fig:models}, emphasizing how they can be used to describe human information networks.

}

\hlc{



All models based only on time and topology (See Figs.~\ref{fig:models}\emph{a-e}) do not include information about messages, documents or text. A simple extension adding a text attribute to the edges would still be less expressive than our model, because this simpler solution would not be able to differentiate between different types of communication such as unicast, multicast and broadcast. These are instead allowed in our model exploiting the presence of nodes representing text messages, and thus justifying the adoption of a bipartite model instead of the simple graphs used in contact sequences. Single time annotations are also unable to distinguish between production/consumption or sending/receiving time. In summary, contact sequence models (Fig~\ref{fig:models}\emph{a}) can be expressed using our model by representing edges as edge-message-edge triples, but contact sequences cannot represent all the information that we can express using our model.
Time-slices (Fig~\ref{fig:models}\emph{b}) and longitudinal models (Fig~\ref{fig:models}\emph{c})
can also be obtained starting from our model, as we do not make any assumption about how the time is represented on the edges. It is thus possible to represent both time-slices and longitudinal models as temporal text networks by just creating a new message $m_j$ and a sequence of edges $(v_i, m_j, l), (m_j, v_k, l)$ for each original edge $e = (v_i, v_k, l)$ in the layer $l$ of the sliced network.
Finally, when only time and structure are concerned, memory models (Figs.~\ref{fig:models}\emph{d-e}) are usually constructed from contact sequence models by aggregating the edges conditional on preceding pathways. While the original temporal information is partially preserved during the creation of the memory model, it is impossible to preserve more information from our temporal text network such as messages or network attributes. 
Therefore, we can think of our model as a way to represent raw and complete information about the temporal interactions and memory models as a way to emphasize information provenance. However, to represent provenance we need to allow edges between messages, and for this reason only our extended temporal text network model is able to express all the information present in memory models (in addition to text, multicasting and production/consumption times, as for all the other models not based on bipartite graphs).

}

\hlc{


The absence of relations makes it difficult to describe human information networks using just time and text (Figs.~\ref{fig:models}\emph{f-g}), 
and despite their versatility to analyze text documents, strictly speaking none of the models only focusing on text and topology without actors (Figs.~\ref{fig:models}\emph{h-j}) allows us to represent human-information networks as they do not contain any representation of the consumers and producers of the text. When also actors are represented, as in some HIN-based models (Fig~\ref{fig:models}\emph{k}) and in socio-semantic networks (Figs.~\ref{fig:models}\emph{l-m}), our model adds directionality, which is necessary to represent text sender/receiver and producer/consumer relationships. Time is also not typically used in these models, but a temporal extension of existing HIN-based and basic socio-semantic models is straightforward and has in fact already appeared in the literature (Fig~\ref{fig:models}\emph{m}). The application of socio-semantic networks are also limited if compared with our model, as they contain already processed information (concepts) rather than text. With this we do not mean that our model is superior, as it can be useful to process the text into concepts, but this shows how we can go from our model to a socio-semantic model but not the other way round.





}

\hlc{

Citation networks and author-citation networks (Figs. \ref{fig:models}\emph{n-o}) can represent relationships between messages, and thus require our extended model to express their information. However, they cannot express communication, because even the more expressive author-citation network model (Fig~\ref{fig:models}\emph{o}) only focuses on the production of text. In particular, there are no edges between documents and authors, but only (implicit) edges between authors end documents. Spreading processes (Fig~\ref{fig:models}\emph{p}) also share the same limitations of either contact sequences or author-citation networks, depending on whether messages and/or authors are represented in the specific model, in addition of not (typically) keeping the text content, which is however a minor problem as text can be easily added to the nodes representing the shared items.
Compared with our core model, polyadic conversations (Fig~\ref{fig:models}\emph{q}) can express almost the same information: both can express unicast, multicast and broadcast relations between messages and actors, both differentiate between information producers and consumers and contain the raw textual information. However, while in our model each individual edge connecting messages and consumers can have a different temporal attribute, in the polyadic conversation model each polyadic interaction has one single temporal value.

}

\section{Analyzing temporal text networks}
\label{sec:analysis}

One reason to adopt a common model instead of defining ad hoc models for each application is to reuse existing analysis methods. While our model can be analyzed \emph{directly}, for example studying dynamical processes such as text propagation in a similar way as in our motivating example, we can consider other strategies. Here we define two more approaches that can be used to analyze temporal text networks: we call them \emph{continuous} and \emph{discrete}.

The practical benefit of using these two approaches is that instead of developing new algorithms the analyst can focus on defining mapping functions encoding the model in a way that fits the data and analysis at hand. Then, these functions automatically generate model views of which existing algorithms can be computed.

\subsection{Continuous analysis}
\label{subsec:continuous}

The main idea behind this approach is to map the elements of the network (e.g., actors, messages, content, etc.) into an asymmetric metric space. This means that it is possible to compute distances between them.

Once distances are available, one can directly reuse existing data analysis methods for metric spaces, such as traditional distance-based and density-based algorithms (k-means, db-scan, etc.). Distances can also be used to retrieve relevant information from large temporal text networks, specifying an information query as an element of the metric space and retrieving those elements that are the closest. We present an example of this last type of analysis in the next section.

\begin{figure}[th!]
\centering
  \includegraphics[width=0.45\textwidth]{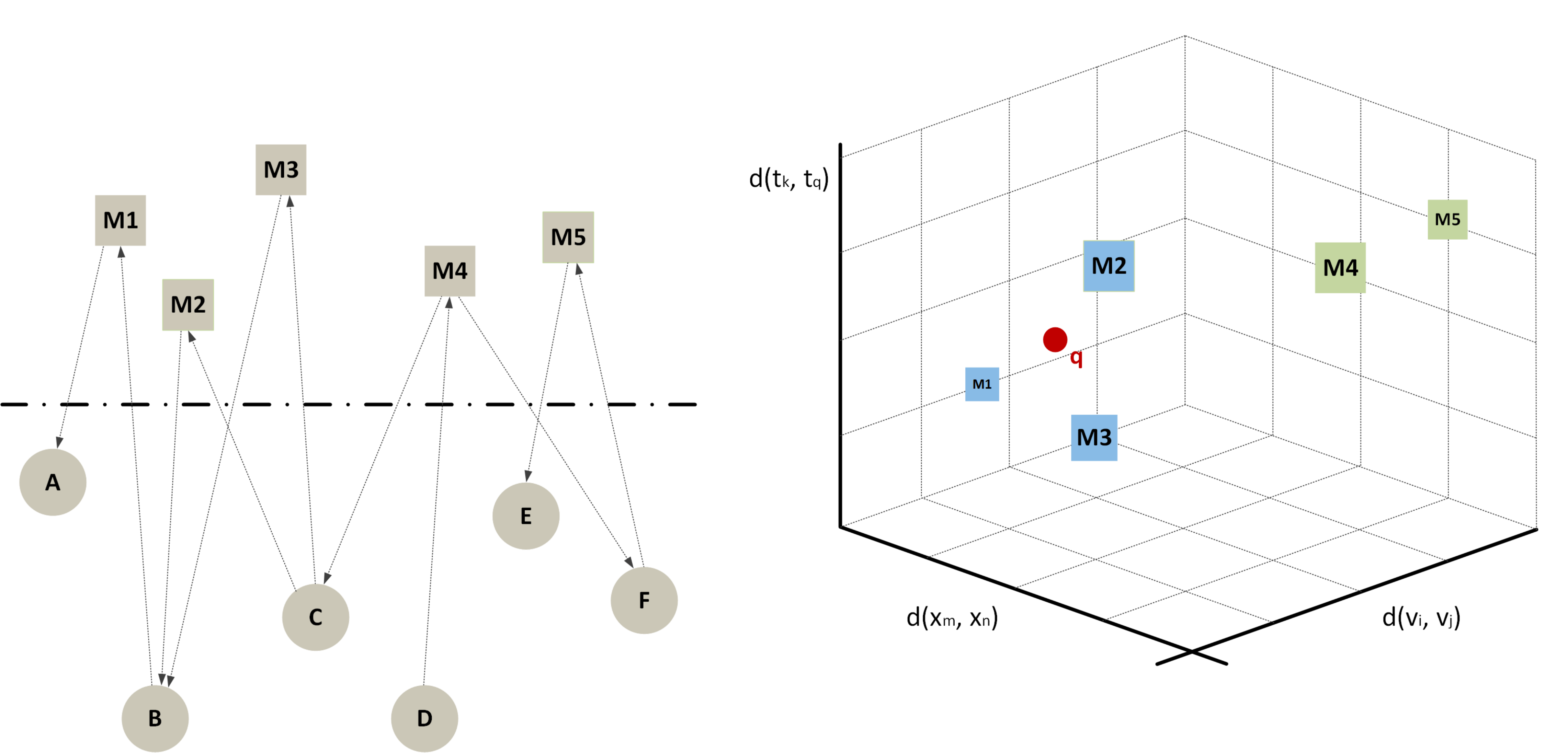}%
  \caption{\textbf{Continuous approach: embedding}. (\textit{left}) A temporal text network with 6 actors --- circles --- and 5 messages --- squares; (\textit{right}) the messages have been grouped into two clusters based on their topological, temporal and textual distance. The point marked with $q$ represents a user's information requirements; in this example the left cluster $(m_1, m_2, m_3)$ contains nodes that are more relevant for the user.}
  \label{fig:continuous}
\end{figure}

The first way of doing this is to use a network embedding method \cite{GoyalF17embedding}. While network embedding was initially defined for simple graphs, more recent algorithms can be directly applied to attributed graphs \cite{Huang:2017:LIA:3018661.3018667}. Meanwhile, we foresee the definition of special versions of these algorithms that are specific for temporal text networks. Figure~\ref{fig:continuous} shows an example of this first type of translation, where messages are the target of the analysis. The same approach can also be used to study other structures and elements in the temporal text network such as actors or combinations of actors and messages.

\begin{figure}[th!]
\centering
  \includegraphics[width=0.45\textwidth]{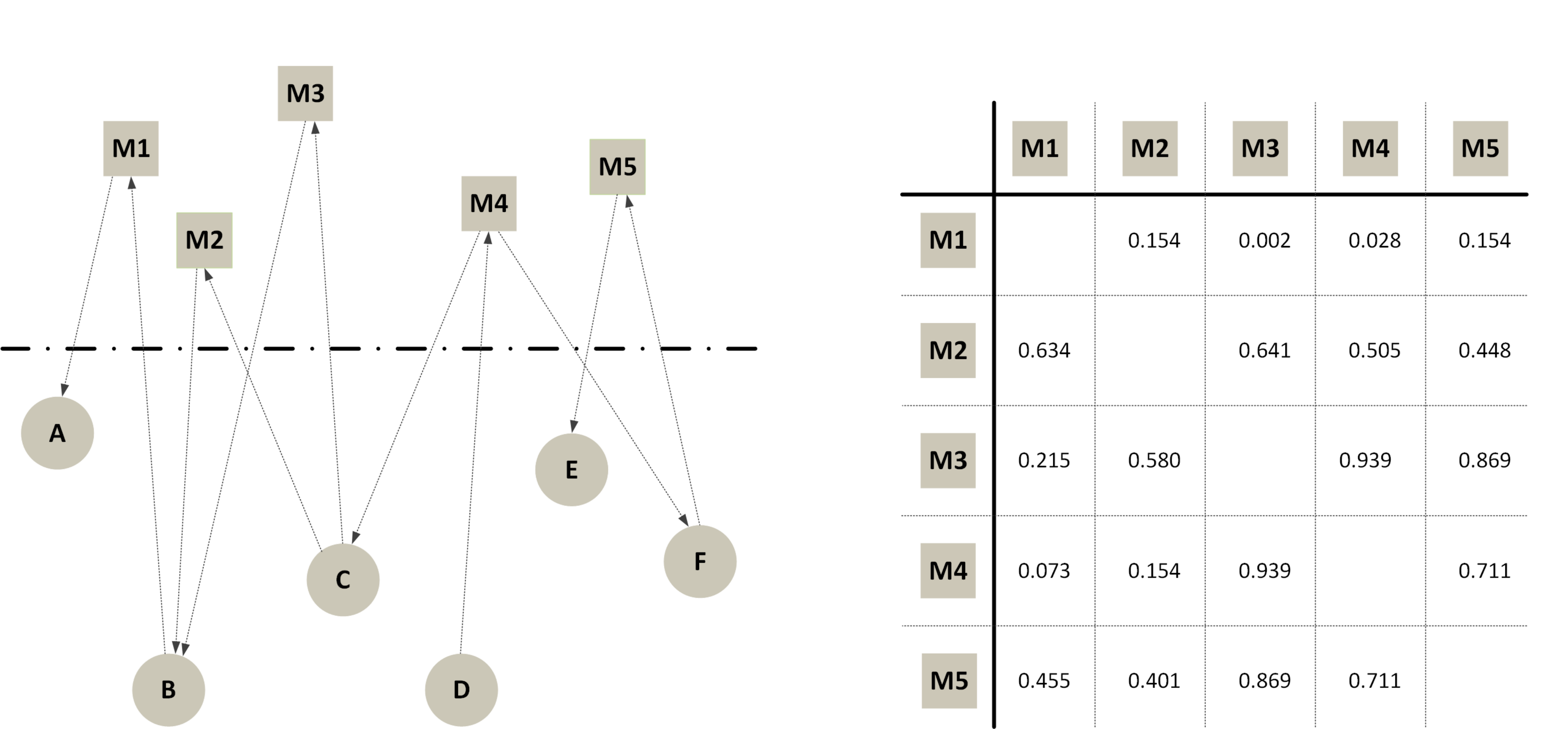}%
  \caption{\textbf{Continuous approach: distance-based}. (\textit{left}) A temporal text network with 6 actors --- circles --- and 5 messages --- squares; (\textit{right}) a messages' distance matrix is obtained from the network \hlc{topology and time attributes}.}
  \label{fig:continuous:distance}
\end{figure}

The second way to use the continuous approach is to directly define a distance function, without any explicit embedding into a coordinate system, so that the points form a metric space but have not an explicit position: only their relationships are defined. This approach is represented in Figure~\ref{fig:continuous:distance}.

The two approaches may look similar: in both cases algorithms use distances, which can be computed after an embedding or are directly defined in the distance matrix. In practice, however, there can be relevant differences. For example, after embedding it is easier to index the data so that not all distances must be computed when algorithms are executed, leading to lower computation time. On the other hand, the direct usage of a distance function is more natural if distances are asymmetric, e.g., when $d(M1,M2) \neq d(M2,M1)$. Asymmetric distances often appear in temporal and directed networks, that are both features of our model.

\subsection{Discrete analysis}
\label{subsec:discrete}

The main idea behind this approach is to encode temporal and textual information into network structures, in particular layers in a multilayer network, so that methods from multilayer network analysis can be directly applied \cite{Kivela2014,DickisonMagnaniRossi2016}. This can be done by defining a mapping function from time and text into a discrete set of classes that are relevant for the analysis. Then, topic-and-time-based user centrality, topic-and-time-based relevance, as well as community detection algorithms can be used. An example of this last type of analysis on real data follows in the next section.


\emph{Textual discretization} is typically performed using
methods from Natural Language Processing such as topic, sentiment or semantic analysis. The main objective of the procedure is to group together messages whose contents have similar characteristics.
\emph{Time discretization} is apparently simpler, because only the cutting points between time slices must be indicated. However, also time discretization presents many options. First, there are often many ways of defining the cutting points, leading to different results. Second, after the cutting points have been defined there can still be different ways of distributing network structures into the slices. For example, if we want to discretize messages, 
we can place a message $m_i$ in a specific interval $(t_a, t_b)$ either if the incoming edge $e = (v_j, m_i, t)$ exists in the interval $(t_a, t_b)$, if all the edges from/to $m_i$ exist in the interval, if at least one of the out-going edges $e = (m_i, v_j, t)$ exist in the interval, etc.
Finally, we use the term \emph{multiple discretization} when both textual and time discretization are applied together to generate the different groups.

\begin{figure}[th!]
\centering
  \includegraphics[width=0.45\textwidth]{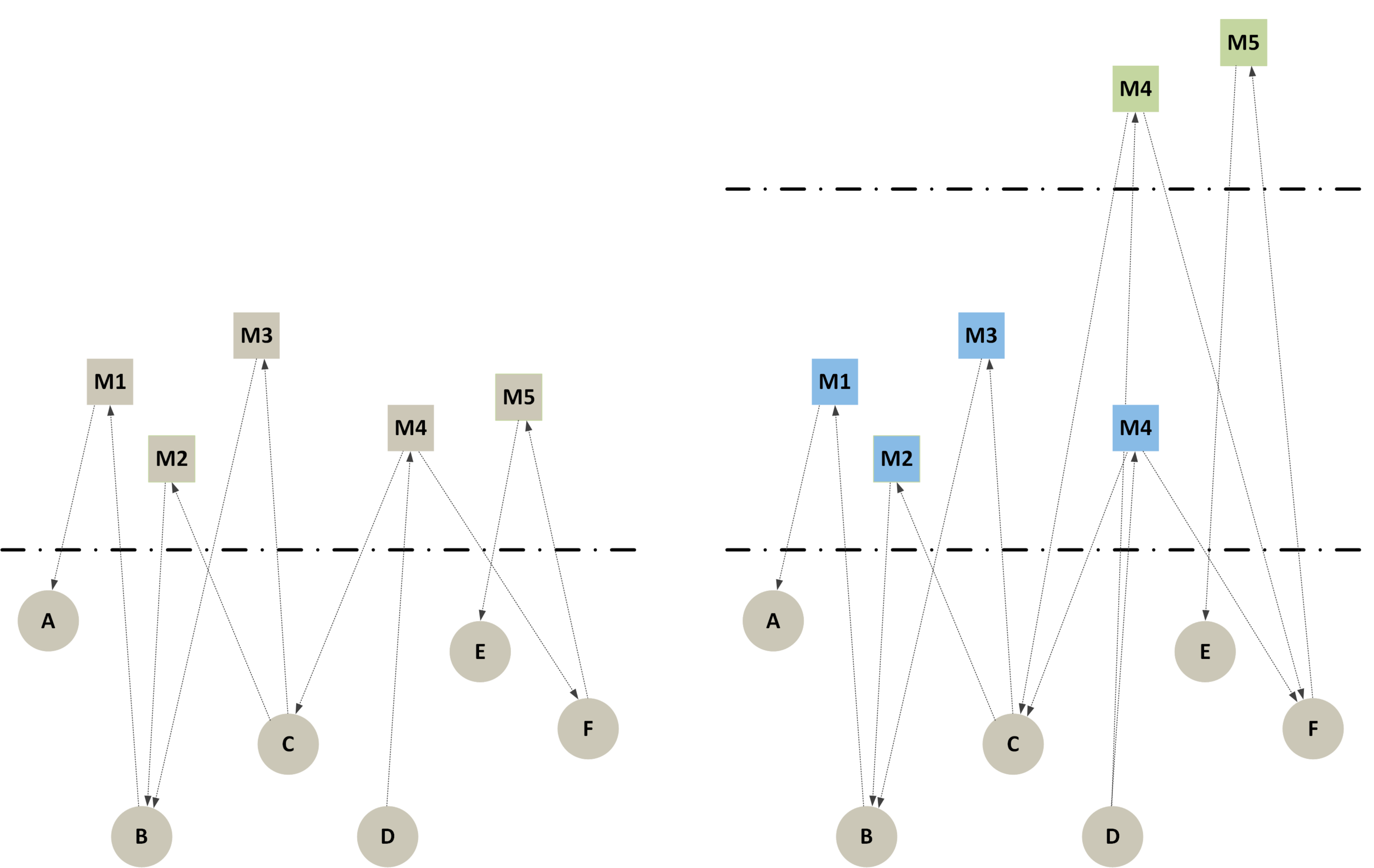}%
  \caption{\textbf{Textual discretization}. (\textit{left}) A temporal text network with 6 actors --- circles --- and 5 messages --- squares; (\textit{right}) the network has been discretized into two clusters --- the top one with 2 messages, the bottom one with 4 --- based on the topic of the messages.}
  \label{fig:discretization}
\end{figure}

Under this procedure, our model would produce a k-partite network with one partition for each new cluster of messages and one partition for the actors. The procedure to generate such network is straightforward once the discretization function is defined. 
Figure~\ref{fig:discretization} shows an example of textual discretization 
where the resulting 3-partite network contains the original layer of actors $A$, and two message layers with 2 and 4 messages each grouping together messages about the same topic. In this particular example, $x(M_4)$ was related to both topics, therefore the message $M_4$ appears in both layers. \hlc{A similar network structure will emerge from time discretization.}



\begin{figure}[th!]
\centering
  \includegraphics[width=0.45\textwidth]{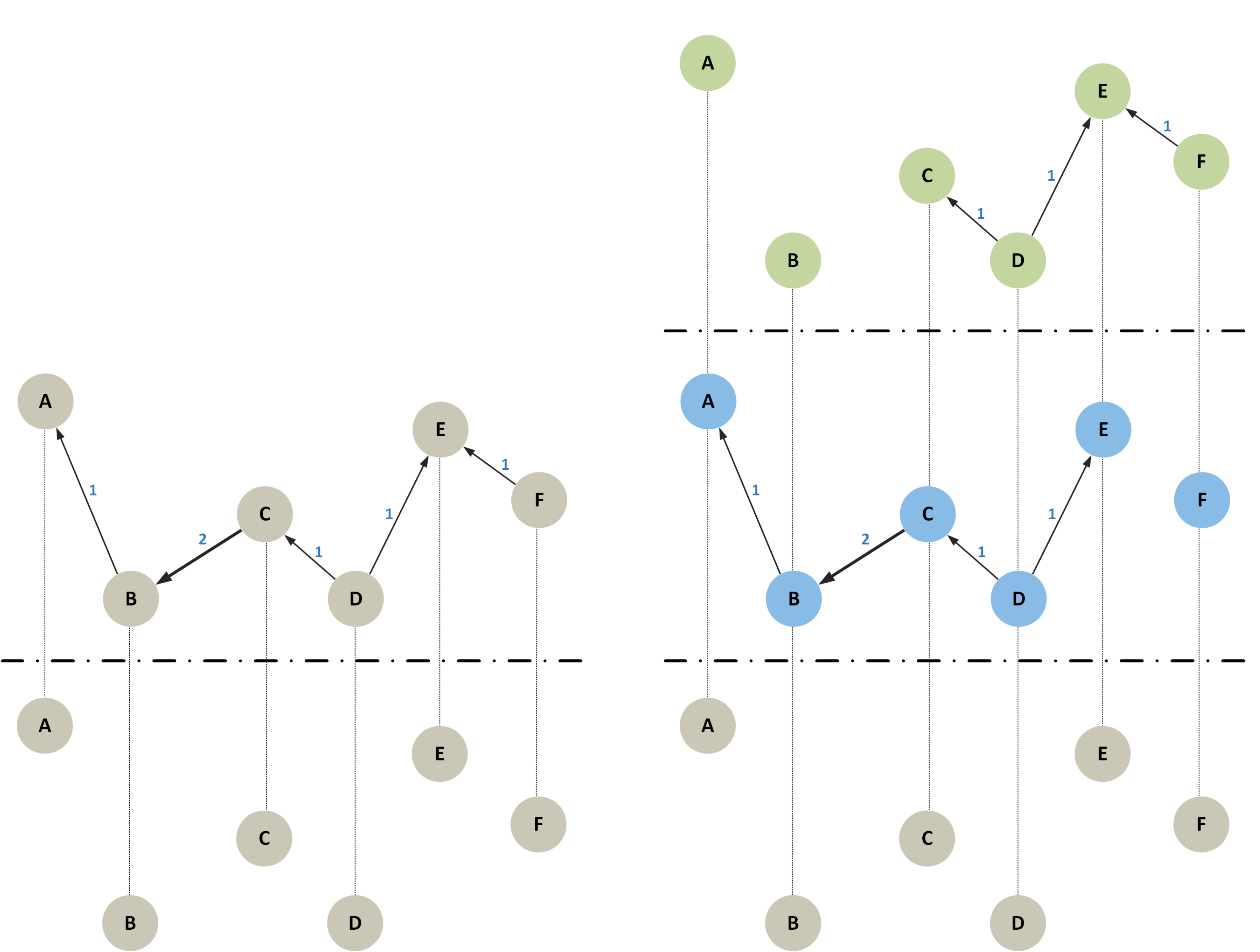}%
  \caption{\textbf{Projection}. (\textit{left}) A projection of the message layers into the actor layer in the original bipartite network in Figure~\ref{fig:discretization}-\emph{left}. The projected multilayer network has 6 actors, 12 nodes and 5 weighted edges; (\textit{right}) a similar projection using the 3-partite network described in Figure~\ref{fig:discretization}-\emph{right} which generates a multilayer network with 6 actors, 18 nodes and 7 weighted edges.}
  \label{fig:projection}
\end{figure}

An additional operation on multilayer networks that can be applied to the discretized data is projection, 
creating edges in one layer based on the information present in another layer. In the resulting multilayer network, a new edge $e_{ij}^{[l]} = (v_i, v_j)$ is created if there is a message $m_k$ in the partition $l \in L$ of the original network with: a) an edge $(v_i, m_k)$  from actor $v_i$ to message $m_k$ and b) an edge $(m_k, v_j)$ from message $m_k$ to actor $v_j$. Weights can also be added to the new edges, using various methods. Figure~\ref{fig:projection} shows one possible projection from the network in Figure~\ref{fig:discretization}. 
In this example the content of the messages (and more in general also the time) are now encoded into the relations between actors.

The main advantage of using a projected multilayer network to analyze temporal text networks is the vast available literature that has targeted this type of data.
In Section~\ref{subsec:discrete_analysis} we use the approach described above together with a clustering algorithm for multilayer networks 
to find communities of actors discussing about the same topics during the same time spans.

\section{A case study}
\label{sec:case}

In this section, we apply the model and approaches introduced in Sections~\ref{sec:model} and~\ref{sec:analysis} to a real temporal text network. 
In particular, we focus on 
\hlc{using} the discretization approach introduced in Section~\ref{subsec:discrete} to analyze the formation and evolution of communities of actors and messages.

The objective of this section is two-fold. First, we want to give \hlc{a} concrete example of the abstract type of analysis described in the previous section\hlc{.} 
Second, we want to show in practice how a new type of analysis can be easily built as a composition of the \hlc{transformations} introduced in the previous section and an existing algorithm (Section
\ref{subsec:discrete_analysis}).

\subsection{Dataset}
\label{subsec:dataset}

Our initial dataset consists of 247,399 public tweets with the hashtag \emph{\#iot} (Internet of Things) or some of its variants (e.g., \#IoT, \#IOT, etc.) automatically collected using the Twitter streaming API in June, 2017. The dataset contains mentions (tweets including \emph{@username}), retweets (tweets starting with \emph{RT @username}), other tweets that are neither mentions nor retweets, and the 51,369 users involved in the aforementioned communications. 
In order to improve the homogeneity of the collected data we further filtered our dataset by keeping only the tweets using at least one of thirty-two hashtags 
selected by domain experts as representative of main topics in this domain. This operation removed for example tweets containing the string \emph{\#iot} but not concerning the Internet of Things.
In the following experiments we focus on the network obtained starting from the tweets containing mentions (about 5\% of the initial tweets), built by coding each tweet as in Figure~\ref{fig:twitter_ttn}. 

\begin{table*}[!h]
  \centering
  \small
  \caption{\textbf{Temporal text networks} used in the case study and its basic properties: number of actors ($|A|$), number of messages ($|M|$), number of edges ($|E|$) and number of layers ($|L|$).}
  \label{tab:datasets}
  \begin{tabular}{lccccc}
    \toprule
    {\bf Networks} & Type & $|A|$ & $|M|$ & $|E|$ & $|L|$ \\
    
    \cmidrule[0.4pt](r{0.125em}){1-6}%
    
    \emph{Original} & bipartite & 15,717 & 13,210 & 35,015 & 2  \\
    \emph{Discretized} & k-partite & 15,717 & 17,273 & 44,943 & 182  \\
	\emph{Projected} & multilayer & 15,717 & - & 23,766 & 182  \\ \bottomrule
  \end{tabular}
\end{table*}

The resulting temporal text network contains about one third of the users in the initial dataset (15,717) and the 13,210 messages exchanged between them (See Table~\ref{tab:datasets}). We call this the \emph{original network}, and use it as a the starting point for both the following experiments.

\subsection{Discrete analysis}
\label{subsec:discrete_analysis}

Social interactions within a group of participants can form a community if they occur more frequently within the group than with other members of the network. In temporal text networks, those interactions are the result of the exchange of messages between actors. In this example we show how our model can be used to find communities of actors discussing about the same topics during the same weeks. Following the method described in Section~\ref{subsec:discrete} we first transform our network to a multilayer network preserving information about interactions between users, topics and time, so that we can then apply an existing clustering algorithm.



The \emph{discretized k-partite network} is built following the procedure explained in Section~\ref{subsec:discrete}. In this particular example, we first split the original layer of messages using their hashtags as an indication of the topic, then we further discretize based on the week when messages are posted. 
The second discretization uses the posting time to create hashtag-week-specific layers. 


Finally, we build the \emph{multilayer network} by projecting each one of the layers containing messages into the actors' layer. Two actors in this network are connected in a given layer $L = (h, w)$ if at least one of them has sent a message to the other using the hashtag $h$ during the week $w$. If multiple messages have been exchanged between two actors in the same layer, only a single edge is generated during the projection. At this step all edges are undirected and unweighted to fit the community detection algorithm we used. Table~\ref{tab:datasets} describes the main properties of the original temporal text network, the projected k-partite and the final multilayer network used during the analysis.

Using the multilayer network and the clique percolation mechanism described in~\cite{Mucha876}, we proceed to detect communities of actors across the whole network. 

Figure~\ref{fig:com} shows the communities with more than 3 actors formed in the multilayer network. Communities contain users and topics, and both users and topics can overlap across communities. The number of users is indicated by the size of the community, while the layers representing the topics of interest of the actors are annotated next to each community. The smallest community in the diagram has 4 actors in the same layer, while the largest community contains 27 different actors and 3 layers. The edges between communities in different weeks indicate that at least one third of the users in the second community were also present in its predecessor. The thicker the line, the more users are shared between them. 
 
\begin{figure}[h!]
\centering
  \includegraphics[width=0.45\textwidth]{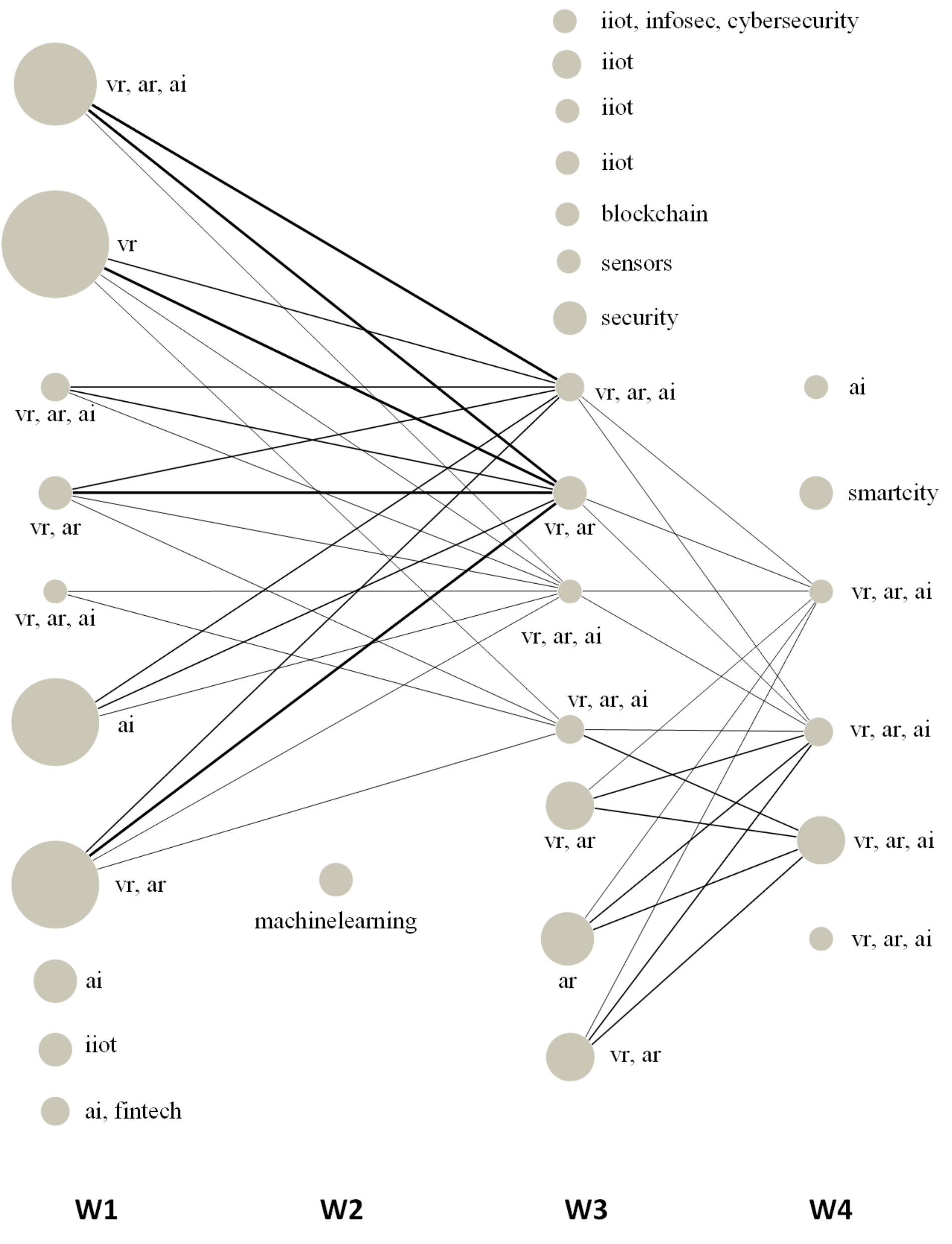}%
  \caption{\textbf{Evolution of communities in the IoT space}. The size of the communities is indicated by the size of the nodes --- representing the number of actors --- and the annotated hashtags. The thickness of the edges between two communities indicates the number of common actors between them.}
  \label{fig:com2}
\end{figure}

We can observe that some of the hashtags, in particular \emph{artificial intelligence} (\#ai), \emph{augmented reality} (\#ar) and \emph{virtual reality} (\#ai), are very popular in the IoT space, with several groups of interest of different sizes forming around one or more of them. However, while the three topics are present across the whole month, the communities they form are very volatile. Only one of the smallest community with just 4 actors, for example, is preserved in time without changing its members or the topics they discuss. The largest communities formed during the first week, instead, disappear in week 2. Later on, some of the same users form new communities but with less members and a higher variance of topics.
Less frequent hashtags such as \#machinelearning, \#security, \#sensors, \#smartcity and \#blockchain also form groups of interest, usually smaller and with no or a few connections with the groups of users discussing the most common topics. 

Overall these results suggest that the IoT space is very fragmented in this Twitter dataset. None of the found communities was big enough to become the main arena to develop a long-standing conversation on a specific topic. Instead, users organize themselves in smaller groups that change over time. Without combining topology, text and time we would find bigger  communities, that would however include users talking about different things and at different times.

In summary, this example shows how a new analysis method can be easily constructed using our model and the approaches described in the previous section. In addition, also the results of this experiment highlight the value of using all the elements of the temporal text network in the analysis.

\section{Discussion and conclusions}


In this work 
we introduce a general model to represent temporal text networks based on the principles of expressiveness, simplicity and tractability. Our model is expressive and simple enough to encode the key components of human information networks (topology, time and text) into a single bipartite network, so that we can represent a range of different forms of communication and data sources spanning from postal services to online social media. 

We additionally show how the model can be analyzed either directly or indirectly, to perform a variety of mining tasks. In particular, we define various transformations for two approaches that we call continuous and discrete. Using such transformations, we can map the data into existing models, allowing to reuse part of the machinery already developed to analyze complex data. While we do not describe each one of the possibilities enabled by our model in detail, in the experimental section we show two concrete experiments using the aforementioned transformations to analyze a set of communication messages exchanged in the Twitter platform during 
June 2017.  


During the past century, the research community has demonstrated a huge interest in studying human information networks. As a consequence, researchers from different disciplines have devoted a considerable time to develop new models and methods to describe aspects of interest in this scenario. However, as we have shown in our review, there has been none or few successful attempts to unify the literature under a common framework: several models and algorithms have been proposed, but only for a subset of the aspects we consider in this article or they have been developed ad hoc to address a specific problem. So, results in one area cannot be directly applied to other types of data. 
We believe that our work can play a key role in the process of consolidating existing efforts from different disciplines under a common framework,
in the establishment of a common terminology and in the 
development of new analytical software able to cope with the complexity of such data.  \\

\textbf{Acknowledgements.} We would like to thank Luca Rossi and Irina Shklovski for the selection of the hashtags used in the experiments.


\bibliographystyle{elsarticle-num} 
\bibliography{bibliography}

\end{document}